# A relationship and not a thing: A relational approach to algorithmic accountability and assessment documentation

**Jacob Metcalf*[1], Emanuel Moss*[1,2], Ranjit Singh*[1], Emnet Tafese[1], Elizabeth Anne Watkins[1]**

Central to a number of scholarly, regulatory, and public conversations about algorithmic accountability is the question of who should have access to documentation that reveals the inner workings, intended function, and anticipated consequences of algorithmic systems, potentially establishing new routes for impacted publics to contest the operations of these systems. Currently, developers largely have a monopoly on information about how their systems actually work and are incentivized to maintain their own ignorance about aspects of how their systems affect the world. Increasingly, legislators, regulators and advocates have turned to assessment documentation in order to address the gap between the public's experience of algorithmic harms and the obligations of developers to document and justify their decisions. Much is at stake in the question of how impact assessment regimes structure the relationships between parties—who must report to whom is as important as what gets reported. In widely used sociological descriptions of how accountability is structured through institutions, an "actor" (i.e., the developer) is accountable to a "forum" (i.e., regulatory agencies) empowered to pass judgements on and demand changes from the actor or enforce sanctions. However, we contend a robust accountability relationship centering on assessment documentation requires a *triadic* relationship, wherein the forum is also accountable to another entity: the public. Typically, as is the case with environmental impact assessments, the public makes demands upon the forum's judgements and procedures through the courts, thereby establishing a minimum standard of due diligence. However, issues of standing and expertise currently prevent publics from cohering around shared interests in preventing and redressing algorithmic harms; as we demonstrate with multiple cases, courts often find computational harms non-cognizable and rarely require developers to address material claims of harm. Constructed with a triadic accountability relationship, algorithmic impact assessment regimes could alter this situation by establishing procedural rights around public access to reporting and documentation. Developing a *relational* approach to accountability, we argue that robust accountability regimes must establish opportunities for publics to cohere around shared experiences and interests, and to contest the outcomes of algorithmic systems that affect their lives. Furthermore, algorithmic accountability policies currently under consideration in many jurisdictions must provide the public with adequate standing and opportunities to access and contest the documentation provided by the actors and the judgments passed by the forum.

CCS CONCEPTS • **Social and professional topics ~ Computing/technology policy ~ Government technology policy ~ Governmental regulations** • Social and professional topics ~ Professional topics ~ Management of computing and information systems ~ System management ~ Technology audits

Contact: jake.metcalf@datsociety.net, em683@cornell.edu, ranjit@datsociety.net, emnet@datsociety.net, ew4582@princeton.edu

* Equal Contribution
[1] Data & Society Research Institute
[2] Cornell Tech Digital Life Initiative
[3] Princeton University Center for Information Technology Policy

## 1 INTRODUCTION

An expanding range of disciplines are concerned with how algorithmic systems might produce harm, and how that harm comes to be recognized and acted upon. Whether organized around the pursuit of 'trustworthy AI' [21,42,46], 'data ethics' [113], 'algorithmic fairness' [30,62,71], or 'responsible innovation' [20,57,93], these practices are focused on identifying, understanding, documenting, mitigating and ultimately avoiding the ways in which algorithmic systems present a danger to individuals, communities, institutions, ecosystems, and society writ large.

Questions around who has knowledge about such harms, and how they come by that knowledge, have also been raised in recent public controversies in computing ethics, such as whistleblower Frances Haugen's claim that Instagram's internal research showed how some of its features harm adolescents [47], Google's firing of Timnit Gebru and Margaret Mitchell after their research critically examined Google's own large NLP modeling [90], and the fraught relationship between Facebook and academic researchers [61]. Amidst these controversies, multiple jurisdictions have proposed regulations and legislation that would require developers to self-study the consequences of their algorithmic systems [27]. It is therefore imperative to consider not only *what* is documented about these systems, but *who* is required to disclose this information and *who* is empowered to receive it and make demands upon the developer to alter the system.

"Algorithmic accountability" has been used as a catch-all term to encapsulate the broad range of responses in the face of these diverse challenges. However, like many social constructs, accountability "is a relationship and not a thing" [95]; it is a matter of how parties in an accountability regime are brought into a relationship with and made responsible to each other. Prior work in this space has spanned across "two concepts of accountability" [17:948], which respectively define it either as 1) a virtuous property of individuals [73] or systems [35,64], or 2) a relationship between differently-positioned parties [69,87,110]. In this paper, we focus on the *relational* nature of accountability to showcase how substantive assessment and documentation practices should be structured to foreground the public interest. This requires taking the *relational* approach seriously in designing and enacting algorithmic accountability regulations to provide footholds for publics to cohere around common interests, have their experiences of algorithmic harms recognized, and seek redress and/or changes to the algorithmic systems.

In conceptualizations of algorithmic accountability, Mark Bovens's studies of organizational accountability have brought significant scholarly attention to the relationship between "actors" and "forums" as the core of accountability structures [69,72,110]. Bovens conceives of organizational accountability as a dyadic relationship between *actors,* and the *forums* that can demand changes from those actors. We build on Bovens' work to propose a *triadic* relationship as the foundation for algorithmic accountability: *actors* (developers or operators of algorithmic systems), *forums* (regulators), and *public(s)* (often achieved formally through jurisprudence). As is the case in other regulatory domains, assessments and documentation can provide the grounds for contestation between these parties, but only when that triad is structured such that the public is able to make demands upon the other parties.

We develop a *relational* approach to accountability for algorithmic systems that emphasizes not what the actors do nor the exact outputs of assessment practices, but rather how interlocking relationships of accountability are constituted between the actor, the forum, and the public (with courts[1] often playing the role of arbiter in contests over harms). In this paper, we (1) explain how accountability is a relational construct, (2) explore barriers the public faces in asserting its role in accountability relationships, when demanding changes to harmful algorithmic systems (particularly with regards to accessing documentation and demonstrating legal standing in courts), and (3) provide a framework for understanding how governance regimes, such as the algorithmic impact assessments currently proposed in multiple jurisdictions, can effectively structure accountability relationships to foreground the public interest.

---

[1] Our core focus in this paper on courts in the United States, except when comparisons with other jurisdictions are generative.

## 2   ACCOUNTABILITY IS A RELATION, NOT A PROPERTY

Discourse about "algorithmic accountability" often suffers from a sort of grammatical illusion. The terms "algorithmic" and "accountability" seem to alternately modify each other, allowing the discourse to slip toward discussions of algorithms that are themselves accountable, i.e., algorithms that hold the property of accountability. While "accountable algorithms" have been specified in a narrow, technical sense [58,64], this slippage implies that merely exposing an account of how an algorithm works satisfies a design parameter for "accountability." However, the underlying rationale of "algorithmic accountability" is that "accountability" modifies "algorithms." That is, developers and operators should be responsive to the people who use or are otherwise affected by their algorithmic systems. If there is no formal method by which people affected by algorithmic systems might demand changes to that system, then "accountability" is not meaningful. In other words, accountability resides in the *relations between* the developers, regulators, and public and its collectivities, not in the algorithmic system nor in the developer's practices alone.

The same grammatical illusion is at play in the accountability literature more generally. As briefly discussed above, two concepts of accountability identified by Bovens [17] from amongst the many usages of the term oscillate between "accountability as a virtue" and "accountability as a mechanism". In the former, accountability inheres in the personal virtue of *being accountable*, and accountability studies focus on "normative issues, on standards for, and the assessment of, the actual and active behavior of public agents" [17:947]. This concept is more strongly associated with accountability discourses in the United States, whereas *accountability as a mechanism* is more likely to frame accountability discourses in Canadian, EU, UK, and Australian contexts [7]. In these contexts, accountability pertains to relationships between actors and forums, and accountability is studied through focusing not on whether actors acted in accountable ways, but whether they "are or can be held accountable *ex post facto* by accountability forms" [17:948].

The multiplicity of meanings associated with accountability illustrates its nature as a contested concept [17,48,63]. A full review of accountability is beyond the scope of this paper, but can be traced back to the Domesday Book of 1066 [50], accumulating meanings across the domains where it is applied. The meanings and the practices associated with accountability have mutually shaped these domains [79,94]. The flexibility and contingency of the term suggest that it can apply to the practices of making algorithmic systems more governable, even if current approaches to algorithmic accountability are not yet sufficient to that task.

We contend that one of the central challenges of constructing robust algorithmic accountability regimes is the middle ground that algorithmic systems occupy between individual and collective frameworks. This creates confusion where to look in efforts to map consequences of these systems. Machine learning models "learn" about a population of people to render predictive decisions about an individual. As Viljeon [101] argues, the economic and technological foundations of machine learning are best understood as fundamentally *relational*: features about one person shape the fates of similarly situated persons, for better and for worse. This relating of individuals to collectives for the purpose of generating predictions creates both utility and risks for machine learning. The origin of algorithmic harms is necessarily located in the collectivities captured in large datasets. However, each instance of harm is often most readily identifiable for individuals, as plaintiffs afforded rights within existing governance frameworks.

For example, take a facial recognition system for policing. If it is developed with racially-biased data, this could lead to higher rates of false arrest for minority demographic groups. So, the origin of harm is in the historical relations underpinning the availability of facial data for training data. But the only available remedy is often through demonstrating a particularized harm: an individual's right to recompense for a false or abusive arrest. There is no legal pathway for a remedy that could demand a wholesale rejection of the biased training data and/or the historical relations in which it is embedded.[2] This causes a short circuit, so to speak, where the only available methods for adjudicating and addressing harms to groups (and to individuals by virtue of their membership in groups) are centered on the interests and rights of individuals (regardless of their group membership). Any effective algorithmic accountability regime needs to bridge this gap and provide the structure for collectivities to also make demands upon developers and regulators.

Given the strict focus of U.S. laws on individual harms, this intertwining of individual and collective outcomes "presents U.S. data governance law with a sociality problem: how can data-governance law account for data production's social effects?" [101:8]. Birhane [13] has similarly argued for a relational ethics approach, noting that both Western ethical traditions and machine learning methods treat idealized rational individuals as the sole locus of analysis and intervention. Birhane draws on non-Western/postcolonial ethical traditions and epistemologies, and certain scientific paradigms from cognitive science and linguistics, to develop a relational approach to machine learning practices and ethical governance. By emphasizing the priority of relations over individuals, Birhane hopes to disrupt the assumption that treats discrete predictions as the very purpose of machine learning, and instead looks for opportunities to treat machine learning applications as always "partially open" in the face of complex and emergent social problems [13:5].

What these authors [13,101], and other philosophical accounts [10,11,109], mean by "relational" is varied, but they are consistently pointing to *relation* as the primary unit of analysis, whether the analysis is ethical, practical, epistemic, or ontological. Barad puts it most simply: "relata do not precede relations" [10:334]. For most consequential phenomena, individual entities (relata) can only be resolved out of a field of relatings, and therefore an account of those phenomena needs to include how this (temporary and partial) resolution is practically accomplished. When it comes to algorithmic accountability, a *relational* approach implies that we—those who want to understand how an algorithmic system affects people's lives and who is responsible when harm is done—should focus on the relations that make possible and sustain accountability. A relational approach pushes against the notion that accountability inheres in technical features of a system, or in documents that satisfy static compliance requirements. It pushes for a policy agenda focused on structuring algorithmic accountability to engender possibilities for publics to emerge, cohere, and assert shared interests to regulators and developers who are obligated to listen.

According to Bovens', accountability is produced within "a relationship between an actor and a forum, in which the actor has an obligation to explain and to justify his or her conduct, the forum can pose questions and pass judgement, and the actor may face consequences" [16:447]. Despite the usefulness of Bovens's articulation for algorithmic contexts [110], questions remain about (1) how the forum itself might be held accountable for its power over actors, (2) what that power consists of, with respect to the operation of algorithmic systems, and (3) how the public can make demands upon the forum regarding the adequacy of its process and robustness of its assessments. We contend that a dyadic relationship between regulator (forum) and developer (actor) is inadequate for the purposes of algorithmic accountability because it cannot foreground the needs and interests of impacted communities and individuals. Such a dyad inevitably results in forms of legal endogeneity [37], wherein the terms of the forum's judgment shift toward that of the actor. In such cases, a "regime that relies on good-faith partnership from the private sector also has strong potential to be undermined by the incentives and institutional logics of the private sector" [87:1].

---

[2] For example, in 2021 the victim of false arrest by the Detroit Police Department, Robert Williams, with the support of the American Civil Liberties Union, filed suit demanding recompense and policy changes about the use of facial recognition by Detroit law enforcement [3,96]. His ability to demand an intervention to the harmful algorithmic system was dependent on him being individually injured by it, not based on his relational status to a history of biases in the production and use of training data. These standing issues are examined at length below.

Extending the forum-actor dyad, we suggest a third entity, *the public and its collectivities* implicated by algorithmic systems, ought to be structurally positioned to hold the forum accountable and ensure the forum's interests do not gradually shift toward that of the actors it is meant to oversee [37]. Within a dyadic model, when actors are themselves tasked with implementing oversight practices demanded by a forum (a common regulatory structure), regulatory goals inevitably tend to shift toward the actor's priorities [87]. In our proposed triadic model, the public serves as a counterweighting third entity, and the public interest is most readily pursued through courts and the establishment of legal precedents requiring an accounting of the public's interests. The actor-forum-public set of relations, therefore, requires that governance regimes include opportunities for the public to litigate the effects of algorithmic systems.

Chief among these practices are *recourse* and *documentation*. The ability of data subjects—people who live with and are "both resources and targets" of algorithmic systems [112:2]—to secure due process of recourse when faced with algorithmic harms is firmly grounded in their access to documentation on the workings of the system and its consequences. Such documentation establishes a route for data subjects and publics to contest the lived impacts of algorithmic systems (which are developed by actors and approved by a forum) through litigation. Recourse and transparent documentation are closely linked in jurisprudence [59], and contestation over outcomes forms the necessary means through which accountability relationships can be intermediated. By providing data subjects access to means to contest the harms they suffer, the courts can become: (1) a forum to provide redress for subjects' injuries; and (2) a backstop that ensures the forum is in turn also accountable for the scope of harms it holds the actors (developers) accountable for. Indeed, one of the more potent consequences of algorithmic governance regimes like algorithmic impact assessments [4,69,72,87,106], which requires developers to study and report on the consequences of their systems, is to create points at which the public can contest the adequacy and accuracy of the developer's claims.

## 3 BUILDING CAPACITY FOR ALGORITHMIC ACCOUNTABILITY

Accountability for algorithmic harms faces structural barriers that can be articulated using a *relational* approach. Collectively, the examples discussed below make a case for shifting the ontology of algorithmic accountability away from individual actors or systems, and towards publics and collectivities. This requires understanding what relations and arrangements of social and political power enable impacted people and communities to demand changes and alter how algorithmic systems impact their lives. In regulatory terms, accountability exists between parties that can exert power and make demands upon each other; such demands include making an appeal, asking for permission, requiring redress when faced with harm, staking a claim to be heard, etc. These relations of accountability cannot exist when this capacity to exert power is skewed towards some parties at the expense of others. In our examples, we explore how these power relations are often stacked against the public in a triadic actors-forum-public relationship.

### 3.1 On Documentation Practices and Incentivized Ignorance

Algorithmic systems are predominantly developed by private companies that enjoy generous intellectual property and trade secrets protections [28]. These protections contribute to the opacity of algorithmic systems [26,107], and also present a barrier for external, third-party actors [82] who wish to understand the workings of algorithmic systems in pursuit of accountability. Although internal documentation practices have grown in recent years [49,70,83], the potential harms developers may be concerned with are not identical to their full set that concern the public interest [69,72]. The result is that accountability for potential harms from algorithmic systems depends on what developers permit to be known about the systems they build, or what can be gleaned from the outputs of the system.

Documentation practices are a key component of a relational accountability framework. Documenting how an algorithmic system is built and the proper conditions for its use can enable affected publics to make demands about the deployment of a system or alter the conditions of its use. However, documentation practices also contain an inherent vulnerability: the developer necessarily must do most or all the documenting, enabling them to choose which features or consequences of the system to document. Barring a voluntary decision to share such documents publicly, insights into relevant characteristics to better understand the relationship between system design and algorithmic harm are unavailable to many of those positioned to intervene and address such harms. Given the tight relationship between transparency and jurisprudence [59], without such documentation the public has little recourse. Conversely, documentation practices that are only concerned with limited technical components of interest to the developer are of little use to the public. Currently, documentation and transparency practices are more closely aligned with the interests of developers in demonstrating that compliance with existing regulations and standards of due diligence than with the interests of the public in making demands about when and how algorithmic systems ought to be deployed.

There are also perverse incentives for companies to remain ignorant of, or to obfuscate, the potential harms their systems may produce, as they can be held liable for failing to act on known or foreseeable dangers their products pose to the public. This same dynamic was central to the difficulty plaintiffs had in demonstrating the harms of tobacco products for smokers: tobacco companies knew about the health risks of tobacco, concealed that they knew, and commissioned skewed scientific research that enabled them to argue otherwise to influence public opinion and reduce plaintiffs' access to legal recourse [2]. Similarly, the recent whistleblower-facilitated release of the "Facebook Papers" confirms long-held suspicions that Facebook has conducted internal research indicating affordances of their Instagram product produced emotional harm for younger users and concealed that they knew about these harms [60]. While commentators have suggested this is producing a "Big Tobacco moment" for Facebook [111], implying a watershed for attempts to bring accountability to the company and its algorithmically-driven products, researchers have also raised concerns that such scrutiny will have a chilling effect on both research within the company and for data-sharing partnerships between the company and external researchers [12,77,85].

Technology companies are often "leery of investigating the ramifications of their profit-seeking strategies" [99] themselves, lest they contribute to the case for greater regulatory scrutiny. This form of agnotology [18,80] can act as a shield, and has long played a role in attempts by a range of corporate actors to sidestep responsibility for their actions by claiming ignorance of the consequences [40,52]. However, in many cases, actors remain answerable for harms regardless of any claims of ignorance [73], particularly under strict liability torts [102], and claims of ignorance as a defense only persist when societal norms allow them to do so.

Regulatory demands for documentation mediate between a public that would benefit from access to information about algorithmic systems and recalcitrant corporate actors who do not independently volunteer such information. As Selbst [87] has pointed out, such regulatory obligations may not necessarily be followed perfectly, but they change how ignorance and knowledge are incentivized within particular governance frameworks and shift organizational culture toward practices that comport with accountability. One of the principle aims of impact assessment is "to get the people who build systems to think methodically about the details and potential impacts of a complex project before its implementation, and therefore head off risks before they become too costly to correct" [87:6]. What is crucial here is that regulation has the capacity to produce *meaningful* changes to the internal practices of organization, not the *mere* performance of regulatory rituals that do not necessarily satisfy the substantive goal of regulation.

### 3.2   Making Algorithmic Harms Cognizable to Courts: The Challenge of Claiming Standing

What might compel a forum to keep making demands of the actor when such demands are called for? The forum needs its own "forum." For regulators of algorithmic systems, this secondary "forum" is ultimately the public(s) with a stake in the workings of such systems and their consequences. John Dewey, in *The Public and its Problems,* points out that "there is too much public" and "too many publics" [34:137] to hold government accountable all the time, but there are occasions when an otherwise "amorphous and unarticulated" [34:131] public organizes itself to express its interests in the face of problems and/or issues that affect them. Such publics are brought together relationally, through the same discursive formations that produce problems that evoke public concern, and inhere in that discourse, prior to any articulation of a specific problem or issue [105]. Perceiving these problems, Dewey argues, often requires specialized expertise, and in acting upon them the public manifests its capacity as a forum to hold the government accountable. While ballots continue to remain a powerful mechanism for such accountability, we focus on another way in which algorithmic publics, constituted around shared interests in the operation and consequences of algorithmic systems, might hold government agencies (forums) accountable for their regulatory practices, and by proxy hold developers (actors) responsible for their diligence: the courts.

Courts often provide a backstop to functions of the administrative state, such as assessment and documentation regimes for complex and potentially harmful systems. For example, major developments or economic policies are often contested on environmental grounds based on procedural rights of opponents to have access to accurate documentation and have the negative case (i.e., do not proceed with the development) heard and considered. The primary consequence of access to courts is not that every decision is contested and adjudicated; rather it is to ensure that procedural rights adequately safeguard substantive rights [104]. Courts are therefore a crucial site for determining the lower threshold of due diligence for algorithmic accountability. Expanding access to courts would therefore likely raise this threshold. However, this is easier said than done; admissibility of contests and claims in court comes with its own conditions.

To begin with, states are generally immune from lawsuits brought by citizens; there are often statutory rules that further limit access to courts when the defendant is a government, such as the ability to sue for violations of procedural rights (e.g., [114]). Furthermore, few computational and informational harms are currently recognized as the kinds of harms that can be remedied by courts. The types of harms claimed by plaintiffs, further, have been largely deemed to "lack standing" because they cannot meet standards of being a concrete and imminent "injury-in-fact." This state of affairs constrains accountability for algorithmic harms, by limiting who can be held answerable for these harms, who can seek redress for injuries they suffer, and how changes to risky or harmful systems can be mandated or incentivized.

"Standing" is the key concept for access: it means to "stand before the court." In order for a plaintiff to have "standing," i.e. for a court to hear their case, plaintiffs seeking redress must demonstrate a reasonable connection between another party's action (or omission of an action) and material harms to their interests. This generally requires demonstrating that 1) the plaintiff is a harmed party, with a claim to a cognizable harm; 2) there is causality, a link between the defendant's action or omission and the harm; 3) the claimed harm is redressable by some action the court is empowered to order. A harm is broadly understood as a wrongful impairment or setback of a person's, entity's, or society's interests, where an interest is any outcome in which one has a stake [45,76]. Philosophically speaking, there are many interests which might be wrongfully impaired—essentially any aspect of life that is valued, and/or one's chances to achieve a desired state—and are therefore potentially harmed. However, only a relatively narrow range of interests constitute a legally actionable harm.

Despite this limited range, there have been judicial innovations in questions of standing. For example, contending with forces of industrialization, questions of standing have become broader over time as the courts displaced "the individual victim [...] by a new kind of claimant — the class member, or the statistical victim [— in adjudicating ...] the potential for injury to populations of indeterminate size and composition" [54:38]. Rule 23 of the Federal Rules of Civil Procedure has played a crucial role in formalizing 'class action' lawsuits since the mid-1960s, and has facilitated the emergence of the 'statistical victim,' represented as a class with common features and interests [44]. These rules have allowed the courts to tackle "aggregated and probabilistic harm that seems always to lurk in modern corporate and industrial systems, no matter what steps have been taken for their control" [54:39]. While these measures allowed the courts to handle harms at the scale of formalized victim populations, they also transformed the courts into administrative agencies that defined the standards for eligibility and recovery for claimants [54]. These provisions of engaging with plaintiffs as statistical victims rather than individuals is useful in contests over algorithmic harms. However, how to aggregate probabilistic harm is still an open question [101] with profound implications for struggles to claim standing and for the public to approach the court.

In the United States, standing in a federal court (where most computational harms are likely to be adjudicated[3]) is determined by interpretations of Article III of the U.S. Constitution, which stipulates access to federal courts depends on establishing injury in fact, causality, and redressability. Additionally, suing the federal government for administrative procedural violations prior to material harm being done—such as conducting an inadequate impact assessment or not providing public hearings and input for major policy changes—is further constrained by the Administrative Procedure Act of 1946 [114]. Currently, standing in federal courts is largely governed by the standard set in *Lujan v. Defenders of Wildlife* [127], in which the Supreme Court ruled that standing requires demonstrating "an invasion of a legally protected interest" that meets two additional criteria: 1) it is "concrete and particularized"; and (2) it is "actual or imminent," not "conjectural or hypothetical" [127]. *Lujan* is especially important to standing for computational harms because it decided who can challenge development plans by the US government based on whether harms were adequately addressed through administrative procedures. A consequence of *Lujan* was to narrow the criteria under which plaintiffs could seek redress, requiring them to demonstrate that the defendant harmed them (or will harm them) in a manner prohibited by a statute and specific to their own interests.

In addition to *Lujan*, standing for computational harms is also controlled by the rulings in *Clapper v. Amnesty International* in 2013 [116], a suit against the federal government claiming surveillance harms, and *Spokeo, Inc. v. Robbins* in 2016 [117], a class-action suit against a personal-records aggregator that published inaccurate information about individuals. The results of these cases established a standard that "fear of harm" and "bare procedural violations" do not rise to the level of "injury-in-fact" necessary to be granted standing. Applying these criteria to computational and informational harms has proven challenging for plaintiffs. Errors, breaches, and even intentional abuse may not result in cognizable harms that the courts see as "actual or imminent" and "concrete and particularized" [91].

Privacy law experts have pointed to *Lujan* [127], *Clapper* [116], *Spokeo* [117], and a host of other cases as a thicket of complicated and sometimes contradictory legal precedents, standards, and taxonomies that limit plaintiffs' capacity to reach standing or redress in privacy cases. For example, Citron and Solove [91] argue that the requirement for harms to be "visceral and vested" (their terms coined to encompass the range of precedents) in order to be cognizable establishes a burden that many privacy harms cannot meet. Much like algorithmic harms, privacy harms can be diffuse, speculative, and downstream, but standing requires material events that may not yet have happened (e.g., identity thieves combining multiple stolen databases) and injudiciously limits plaintiffs' options for recourse. Instead, they advocate for a theory of computational privacy harms that could frame the anxiety caused by privacy breaches as a material harm, and so capture downstream risk in a manner similar to environmental harms. Romberg [84] argues that court precedents have made standing so challenging that they have further blunted the capacity of legislators and regulators to statutorily recognize new categories of rights, particularly around information technologies. Waldman has observed that approaching privacy as a contextual phenomenon, like a lot of scholarship does [9,74], renders it "open to attack as ambiguous" [103:696] in court. Algorithmic harms are similarly approached as deeply contextual, which also makes them open to such attacks.

As these privacy scholars have argued, access to courts for computational harms is both challenging and not a panacea. In the following subsection, we illustrate how similar dynamics effectively limit recourse for algorithmic harms. Notably, there is no opportunity to understand, contest, and/or adjudicate harm—the parties have no effective accountability relationships that would facilitate doing so. The developers rarely address the material claims of the plaintiffs. *The challenge of standing is not just about technical access to courts, it is firstly about whether publics are empowered to be recognized.*

### 3.2.1 Representational Harm

Representational harms occur when algorithmic systems reinforce the political, economic, and/or cultural subordination or denigration of individuals based on their group identity [32], which often has impacts to monetization and representation in, and access to, spaces of public discourse. While there is little incentive for firms to publish details on the use the tools of algorithmic moderation, it is evident that there are representational harms being faced by individuals and groups whose content is subject to these tools [65]. In the case of *Newman v. Google LLC & YouTube LLC* [126], Black content creators associated with the Black Lives Matter movement sued YouTube for flagging their content as inappropriate and placing them under "restricted mode" that blocked minors from viewing their videos and removed the possibility of monetization. The courts could not consider evidence for claimed harms of racial discrimination (dismissed due to lack of evidence showing intentional discrimination), impeded freedom of speech (dismissed due to inability to prove that YouTube can act as a state actor), and false or misleading statements (dismissed for several reasons including YouTubes actions not having to do with concerns with commercial advertising). In a similar case, *Divino Group LLC v Google LLC* [125], a group of LGBTQ+ content creators claimed that their content was being classified unfairly by YouTube. The content creators in these suits claimed that the corporation had unfairly removed and archived their videos, causing them to lose revenue. The case was dismissed on grounds of First Amendment rights and misleading advertising under the Lanham Act. The lack of transparent documentation and available expertise on how flagging algorithms (and human moderators) generated such decisions made it impossible to actually adjudicate their claims regarding harms, resulting in a preemptive loss of standing.

### 3.2.2 Explanatory Harm

Currently, the US court system has no template for bringing a case against an algorithmic decision support system, and to demand an explanation about how decisions were rendered about data subjects. There is such a template in the EU's GDPR, in particular the recognition of data subjects' "right to explanation" in relation to decision-systems which make consequential decisions about them and their life chances. The right to explanation is an attempt to ameliorate explanatory harms, which can be understood as the injustice of having critical decisions made about one's life or life chances, such as in hiring, criminal justice, and medicine, that are opaque and resistant to interrogation. While the "right to explanation" terminology never appears in the GDPR, its requirements enact a statutory mandate obligating firms to provide "meaningful information" about the logic behind data-driven decision-making [88]. Selbst and Powles suggest that the GDPR trends towards "strengthening data protection as a fundamental right" [88:235], creating the conditions for accountability relationships between firms, the government, and the public. Workers in the ride hailing industry recently tested these new relational structures when they took two ride-hailing companies to court to demand that data about their earnings, work assignments, and suspensions be disclosed [86]. They won their case, with the result that Uber and its competitor Ola must make both used data and its logic around its use transparent. The GDPR Articles 12-22 furnish a bulwark against "incentivized ignorance." Requiring firms to provide data subjects insight into how data-driven decision-making is operationalized obligates them to create and maintain adequate records to that effect.

### 3.2.3 Allocative Harm

In the context of algorithmic systems, allocative harms occur in the context of a system tasked with distributing some resource or opportunity, e.g. credit, or jobs [41], where algorithmic systems allocate resources to some social groups more or less favorably than others [15]. In some instances, even where allocational differences fall across protected categories like race, gender, or disability, significant challenges prevent redress for harms through courts [89], despite explicit regulations against discrimination [1]. These challenges include the fact that (potential) plaintiffs may not even be aware an algorithmic system was used to make an allocation decision [81] and that the basis of such a decision may be inscrutable [51]. It is difficult to seek redress for allocational harms unless that harm has a discriminatory dimension, which in turn must pass stringent tests as to whether discrimination has occurred [92]. External audits on hiring algorithms have provided stark illustration of the arbitrary and unjust nature of such systems. For example, a group of German journalists conducted an inquiry into an algorithmic hiring system which used computer vision to assess videos of job interview candidates, conducting A/B tests to see how a subject was "scored" for attributes like conscientiousness, neuroticism, and agreeableness [19]. Disconcertingly, arbitrary changes in a video's visual appearance, such as the addition of a bookcase in the background or a change in the overall tint of the video itself, led to significant changes in the scores granted by the system to job candidates. Such audits, while critical to call attention to these systems' issues, cannot alone enable accountability between parties. An individual who was algorithmically excluded from the opportunity to interview for a job would not necessarily know this choice was made by an algorithmic system, and a plaintiff in such a case would need to demonstrate that they had been deprived of an opportunity based on a flawed algorithmic product. Demonstrating that the product was flawed would requires expertise and careful inspection of the algorithmic model, the data on which it was trained, the accuracy metrics for its predictions, and the basis on which it makes inferences about the suitability of candidates. Often, the basis for inferences is opaque even to the developers [26,75], which makes it impossible to demonstrate that a harm has occurred because of an algorithmic inference.

### 3.3   On Expertise and Consensus in the Adversarial Process of Court Proceedings

A courtroom is a site for reenacting a sequence of events that led to a contest of accountability requiring adjudication: "someone must be blamed, someone punished, someone rewarded for exceptional enterprise, someone, if possible, made whole" [55]. In the adversarial setting of U.S. courts, there are competing interpretations of this sequence of events and the associated evidence to establish their veracity. Cross-examination becomes a means to establish ground truth and facticity. Expert testimony is a key component of this cross-examination and provides the groundwork of defining standards and techniques to measure the contested harms as claimed by the plaintiffs. For example, in challenges to environmental impact assessment (EIA) documents produced by developers and approved by the Environmental Protection Agency (EPA), plaintiffs can take EPA to court by challenging: (1) the scope of the assessment, by arguing that it ignored significant impacts that affect them [121,122]; or (2) the adequacy of the assessment, by arguing that the methods used to evaluate an impact inadequately assess the severity of a potential harm [120,124]; or (3) the decision to permit the project, by arguing that the permitting agency ought to have recommended an alternative, less harmful, alternative design be chosen instead [119,122–124]. Making any of these arguments requires expertise on environmental concerns that the plaintiffs must access and bring to bear upon their claims. The kind of challenges raised by plaintiffs also showcase a peculiar aspect of environmental litigations that they are often decided not by ascertaining ground truth of causality and consequences, but rather through cross-examination over whether "scientific assessment procedures were properly followed, including those for soliciting expert advice and subjecting it to the scrutiny required by applicable statutory mandates" [55].

   It is difficult for courts to ascertain the true state of play in any technoscientific field, including computational sciences. Credibility of the witness, in such cases, becomes a crucial resource for evaluating claims. During court proceedings, even before the testimony is presented, the judge must decide in advance whether an expert witness has the right credentials to testify. The Daubert standard (instituted by the U.S. Supreme Court in *Daubert v. Merell Dow Pharmaceuticals* in 1993 [115]) provides the foundational heuristics for evaluating the validity of expert claims according to a set of criteria: "(1) whether the theory or technique underlying the evidence has been tested and is falsifiable; (2) whether it has been peer reviewed; (3) the technique's error rate, if known; and (4) general acceptance" [53:63] of the technique in the particular technoscientific field in which it belongs. The standards of peer review and general acceptance highlight that the courts have paid attention to the social process of building consensus over techniques, standards, and technoscientific facts and incorporated it into determining whether the technique or an expert claim based on generally-agreed-upon facts in a field of technoscientific study is admissible as evidence or not.

Debates over validity of technical as well as qualitative methods used to measure and document the efficacy and consequences of algorithmic systems are currently ongoing [see, for example, the debates over suitability of different methods to measure the impact of the Alleghany Family Screening Tool in 72:45–46]. Standards of peer review and general acceptance, amid such contestations, can become a significant barrier to admissibility of any kind of expert evidence contesting or even relying on algorithmic systems. For example, the standard of general acceptance recently came into play when prosecutors in a murder case withdrew evidence gathered using ShotSpotter—an algorithmic system that relies on microphone sensors to detect and geolocate gunshots—against a defendant [43]. This evidence was withdrawn to seemingly avoid the possibility that ongoing controversies over whether ShotSpotter is accurate in identifying gunshots may persuade a judge to rule against its forensic validity and set a precedent for taking the tool off the table for both future criminal investigations and prosecutions.

Similarly, in *Newman v. Google LLC & YouTube LLC* [126] discussed above in Section 3.2.1, the plaintiffs contested the use of content filtering algorithm by YouTube (which is owned by Google's parent company Alphabet) on the grounds of racial discrimination. Among their evidence was the argument that, "in December of 2020, Google fired Timnit Gebru […], the co-leader of Google's Ethical A.I. team, because Gebru complained about [Google's] 'biased filtering and blocking tools'" [126]. In bringing this event to the Court's attention, the plaintiffs were relying on Gebru's expertise in evaluating bias and discrimination in NLP algorithms used at Google as evidence of racial bias in YouTube's content moderation algorithms. In rejecting this evidence, the court reasoned that it "has no way of knowing if the filtering and blocking tools in question were used only at Google, or if they were also used at YouTube" [126]. The court's reasoning was that while algorithms used by Google and YouTube may be owned by the same parent company and share a family resemblance, they cannot be demonstrated to utilize the same code nor moderation rules. Under current regulatory conditions, YouTube has few incentives and is under no obligation to provide documentation on its content moderation algorithms or assess the consequences of their moderation policies and algorithmic sorting processes in a manner that publics can contest them. Regardless of the merit of the claims made by the plaintiffs, this case highlights how the lack of available documentation and recognized expertise that can be brought to bear upon an algorithmic system makes it harder for a public to cohere around their potential harms, and demand redress.

More broadly, in adjudicating legitimacy and admissibility of expert claims as evidence in courts, judges have come to play the role of gatekeepers to publics [53,55]. Such gatekeeping has had its own exclusionary effects as to which consequences and interests the public can contest in courts, and has tended to favor corporate defendants [38,39]. Despite such gatekeeping, courts remain a crucial backstop for the public to hold the government and by extension, private actors accountable.

## 4 IMPACT ASSESSMENTS AND THE LEGIBILITY OF HARMS

The question at hand for algorithmic governance is whether the accountability structures, documentation, and relationships being considered by governments provide footholds for publics with common interests to emerge and have their voices heard in the algorithmic accountability regime. As legislation and regulatory efforts increasingly consider requiring developers to conduct assessments of their systems, a relational approach to algorithmic accountability provides some guidance around how to foreground the public interest. As discussed above, there are always many possible publics, and it is only through a contingent process that these publics cohere together to speak for their interests [34]. Establishing the right to contest the outcomes of algorithmic systems is integral to this process [59]. Sometimes, these publics are temporary and tactically opportunistic, other times they are robust and sustained; such publics cannot be pre-figured and may cohere around concerns with an emergent technology that are not yet envisioned.

Efforts to cohere publics around algorithmic systems have drawn on regulation and case law for environmental harms [87], public nuisance [8], and regulation of food, drugs, and cosmetics [98] as suitable precedents, given the commonalities of uncertainty in establishing causes and consequences between them. Along these lines, algorithmic impact assessments (AIAs) are increasingly invoked as a mechanism for publics to make demands upon developers [4,69,72,87,106]. Impact assessment regimes are often the appropriate governance mechanism where there are diffuse, probabilistic, and potentially novel forms of harms and many intersecting interests implicated in a proposed development or a deployed system [87]. Inferring from other domains of impact assessments, AIAs have the potential to partially address the challenges of building capacity for algorithmic accountability, particularly issues of legal standing that provide footholds for communities to cohere around procedural and substantive rights and have their interests considered. Where courts now refuse standing for computational harms for not being adequately "concrete" and "particular", or for not demonstrating "injury-in-fact," an AIA regime may create procedural rights that can protect the substantive rights that courts have struggled to recognize. In making algorithmic harms more cognizable to courts, such accountability practices shift the incentives for developers by fostering norms that organizations attend more closely to such harms during development [87] and provide a needed backstop for those harmed or endangered by algorithmic systems to seek due process and remedies.

A central lesson of impact assessment regimes—and environmental impact assessment (EIA) in particular—is that in matters of foregrounding public interests, the *how* of structuring accountability relationships is more important than specifying the *content* of the reporting/documentation process. Under the National Environmental Protection Act of 1969 (NEPA), developers hoping to undertake projects that receive federal funding, cross state lines, or meet other criteria for inclusion, are required to conduct EIAs prior to receiving a permit to proceed with their undertaking. EIAs are expected to thoroughly document a wide range of environmental impacts from the developer's project, including impacts to water quality, wildlife habitat, air quality, cultural resources, soil quality, and so on. In challenging EIAs, plaintiffs contest projects, not by suing the private company responsible for them, but by suing the agency responsible for accepting the EIA conducted to secure a permit. This is a crucial point, as a plaintiff's standing to bring a suit against a private company is much more limited by the requirement that they demonstrate a more direct, material harm. An individual or organization does have standing, however, to bring a suit under NEPA against a government agency "if he or she is adversely affected by an agency action" [14], such as the acceptance of an impact assessment, under the Administrative Procedure Act [114].

The relatively recent evolution of EIAs to include environmental justice concerns [78,100] illustrates the importance of litigating impact assessments. Despite a long history of siting harmful developments in already environmentally-degraded locations populated by disadvantaged communities [22], a growing public sentiment about environmental racism issues [23] and lawsuits claiming environmental racism violated the Administrative Procedure Act (APA) [114], led to an executive order mandating that agencies take environmental justice concerns into consideration [108]. By leveraging procedural rights offered by the EIA process to challenge the completeness of impact assessments that lack an account of environmental justice, a "public" was able to establish a proxy substantive right to have their interests considered, feeding a cycle in which novel or newly-understood harms were studied and then integrated into impact assessments [97]. Developers appealed rulings of regulatory agencies, regulatory agencies tested the boundaries of their authority, community advocates sued regulatory agencies, and new public interest organizations formed to pursue strategic litigation—the accountability relations structured by NEPA enabled publics to emerge, cohere and made demands even if they did not win every ruling.

While AIAs have been implemented only in limited ways (see the use of AIAs in Canada as a self-assessment tool for developers [67]), forthcoming regulatory proposals have been reported as calling for more robust forms of algorithmic impact assessment, although the precise names of the assessment process differ [56]. These more robust forms of AIA would require actors who develop or operate algorithmic decision-making systems to document the expected impacts of such systems, and submit that documentation itself, or a report summarizing the assessment, to a government agency tasked with acting as a forum. In its capacity as a forum, this agency would, among other responsibilities, (1) mandate significant public consultation with stakeholders who might be affected by an algorithmic system; (2) require developers to address harmful impacts that could be ameliorated by changes to the design or deployment of a system; and (3) make aspects of AIA documentation publicly available. As proposed, such regulation satisfies a need for greater understanding of how algorithmic systems produce harmful impacts. Crucially, by locating responsibility for overseeing AIAs within a federal agency, regulatory approaches have the opportunity—if drafted appropriately—to make algorithmic harms more legible to the courts. This also creates conditions for the public, as intermediated by the courts, to be better positioned in holding the forum that oversees the actions of developers accountable.

Suggested proposals for algorithmic accountability have included model cards, datasheets, and a range of first-, second-, and third-party audits [49,70,83]. These proposals depend, to varying degrees, on the prerogatives of those using an accountability tool or conducting an audit; a developer may audit their own product to verify compliance with some narrow requirement or an adversarial third party may interrogate an algorithm to demonstrate it fails to meet a specific set of criteria the auditors are interested in. However, consistently actionable accountability requires demarcated relations of accountability within a community. But audits alone do not a community make, even when they interrogate the issues of greatest concern to a community. Audits are critically important artefacts for bringing attention to problematic systems and have been a key mechanism for the public to assert itself, but they do not by themselves enact a legal obligation with the necessary impetus to compel change in a harmful algorithmic product. Even the consequences of headline-making audits [24,25] have ultimately depended on firms taking voluntary action, motivated by either goodwill or an interest in maintaining public trust that had been threatened by the results of an audit [5,36]. Adversarial audits have become, and will remain, a critical need in the absence of robust accountability relations between actors, forums, and the public. They are particularly critical when a public or community does not or cannot occupy its needful place and stake within a triadic accountability relationship. The rise of adversarial audits may actually be a symptom of the accountability vacuum around algorithmic systems. Algorithmic impact assessment regimes, on the other hand, deliberately create the conditions of possibility for a public to be in relation with both industry and government.

## 5 CONCLUSION

Transparency and documentation can be an appealing solution to a wide range of accountability challenges in the machine learning industry and are familiar governance methods for developers and regulators alike. However, they are not a panacea [6,68]. As we have argued, how the accountability relationships are structured is critical to foregrounding the public interest in regulations for algorithmic systems. However, many of the proposed regulations currently under discussion globally, and including some already deployed, tend to provide little opportunity for public input.

In particular, regulatory structures that create a dyadic relationship between a regulatory agency and developers for reporting and accepting/approving algorithmic impact assessments are at high risk of the regulator becoming fully dependent upon the developer for defining and measuring impacts. It is inevitable that developers will be primarily responsible for measuring their own systems and reporting the outcomes. The question at hand is whether developers alone are permitted to choose the metrics by which their systems are measured or if the public can exert pressure on the thoroughness and adequacy of the assessments.

For example, in the proposed EU AI regulations [31] currently under discussion in the European Parliament, the public regulators define risk tiers, prohibit certain applications deemed contrary to human rights, and require conformance audits for higher risk systems prior to deployment. However, there are few clear opportunities for impacted publics to contest the terms and outcomes of those audits on the basis of shared interests [33]. Similarly, in the Algorithmic Accountability Act of 2019 [29], and the revisions reported to be under consideration [56], rely heavily on the Federal Trade Commission to oversee algorithmic impact assessment reporting, but underspecify under what conditions and by what mechanisms can the public access and contest the reporting.

Such regulations should prioritize establishing a triadic developer-regulator-public accountability relationship that provides clear footholds for the public to cohere around shared interests and contest the impacts of algorithmic systems. In this paper we emphasized the role that access to courts can play in establishing the lower thresholds of due diligence for developers, in large part because courts are currently the primary route *and* primary roadblock for individuals and groups seeking redress for algorithmic harms. Standing in courts, however, is only one (necessary) part of the puzzle and is secondary to the primary goal of providing greater opportunities for publics to cohere around shared interests and make demands for changes to algorithmic systems. We suggest the following features of an algorithmic accountability regime would make the triadic relationship more robust by providing routes for publics to participate in algorithmic accountability:

- *Build feedback loops between regulators and scientific standards bodies.* Standards bodies, such as NIST in the US or the EU Standards Hub, and IEEE and ISO, are an effective route for researchers and advocacy groups to contribute to the formation of measurement practices that developers will use. Regulators are rarely equipped to determine on their own what process developers should follow, and standards bodies are well placed to receive feedback from the public and developers alike to find viable metrics and methods.
- *Specify public rights to review impact assessment documentation.* At a minimum the public should have access to documentation about the purpose and limits of systems, what features it was trained on, criteria used to measure algorithmic fairness, and information necessary to judge the gap between developer's claims about what the system does and its real-world performance.
- *Provide meaningful access to researchers.* Regulators should facilitate researchers seeking to understand broadly how the AI/ML industry is meeting accountability obligations, including providing portals for access to impact assessments and aggregated trends.

**ACKNOWLEDGMENTS**

## 6 HISTORY DATES

Received [month, year]; [review in progress]